\documentclass[12pt,letterpaper]{article}
\usepackage{graphicx}
\usepackage{bm}
\usepackage{amsmath,amssymb,amsfonts,mathrsfs,amsthm}
\usepackage{authblk}
\usepackage{setspace}

\def\x{\mbox{\rm x}}
\def\y{\mbox{\rm y}}
\begin{document}
\date{}

\title{\textbf{Generation of new symmetries from explicit symmetry breaking}}

\author[1]{N. Dimakis\thanks{nsdimakis@gmail.com; nsdimakis@scu.edu.cn}}

\affil[1]{Center for Theoretical Physics, College of Physics, Sichuan University, Chengdu 610064, China}

\maketitle

\begin{abstract}
We study how the explicit symmetry breaking, through a continuous parameter in the Lagrangian, can actually lead to the creation of different types of symmetries. As examples we consider the motion of a relativistic particle in a curved background, where a nonzero mass breaks the symmetry of the conformal algebra of the metric, and the motion in a Bogoslovsky-Finsler space-time, where a Lorentz violation takes place. In the first case, new nonlocal conserved charges emerge in the place of those which were previously generated by the conformal Killing vectors, while in the second, rational in the momenta integrals of motion appear to substitute the linear expressions corresponding to those boosts which fail to be symmetries.
\end{abstract}

\section{Introduction}

Symmetries play an important role in all aspects of theoretical physics \cite{Sundnew}; from the fundamental level and small scales, to cosmology and the description of the universe. The two theorems owed to Noether and her work in the calculous of variations have been most influential in this respect \cite{Noether} (for the English translation by M. A. Tavel see \cite{Tavel}). The first theorem deals with finite groups of transformations, which leave the action of a system form invariant and are related to the well known result of existing conserved quantities, the so-called Noether charges. The second is dedicated to infinite dimensional symmetry groups, which give rise to differential or algebraic identities among the equations of motion, i.e. not all degrees of freedom are independent. These symmetries are related to the existence of some gauge freedom in the respective system. Here, we concentrate on the first type of symmetries and their conservation laws. We are particularly interested to see what happens to the classes of certain symmetries which are broken due to some parameter entering the Lagrangian; this is the scenario of an explicit symmetry breaking.

We start by reviewing a few well known facts from the theory of variational symmetries and symmetries of differential equations. For more details we refer the interested reader to some of the many  textbooks on the subject \cite{Ibra,Stephani,Olver}. For simplicity let us consider a first order Lagrangian $L=L(\tau,q,\dot{q})$ of a system with finite degrees of freedom $q^A=q^A(\tau)$; the dot denotes differentiation with respect to the ``time'' variable of the problem $\tau$. A vector $X$ in the space of independent and dependent variables, $\tau$ and $q^A$ respectively, is given by
\begin{equation} \label{symvecgen}
  X = \chi \frac{\partial}{\partial \tau} + \xi^A \frac{\partial}{\partial q^A},
\end{equation}
and it is called a symmetry of the system if the transformation that it induces leaves the action, $S=\int L d\tau$, form invariant up to a surface term, i.e. $\delta_X(L d\tau)=dF$. In infinitesimal form this is expressed through the well known criterion of invariance \cite{Olver}:
\begin{equation} \label{symcrit}
  \mathrm{pr}^{(1)}X(L) + L \frac{d\chi}{d\tau} = \frac{d F}{d \tau},
\end{equation}
where $F$ is the gauge function related to the surface term and $\mathrm{pr}^{(1)}X$ is the first prolongation of the vector \eqref{symvecgen} given by
\begin{equation}\label{prolong}
  \mathrm{pr}^{(1)}X = X +   \left( \frac{d\xi}{d\tau}^A - \dot{q}^A \frac{d\chi}{d\tau}  \right) \frac{\partial}{\partial \dot{q}^A}.
\end{equation}
That is, the extension of $X$ to include the change in the first derivatives. The components of the vector \eqref{symvecgen}, $\chi$, $\xi^A$ may depend on: $t$ and $q$ (point symmetry), on derivatives of $q$ (higher order or dynamical symmetry) or even possess a nonlocal dependence (nonlocal symmetry).

When we have a vector satisfying the symmetry criterion \eqref{symcrit} we obtain conserved quantities of the form
\begin{equation}\label{genint}
  I = \xi^A \frac{\partial L}{ \partial \dot{q}^A} + \chi \left(L - \dot{q}^A \frac{\partial L}{\partial \dot{q}^A} \right) - F = \xi^A p_A - \chi H -F,
\end{equation}
where in the last equality we substituted the momenta $p_A = \frac{\partial L}{ \partial \dot{q}^A}$ and the Hamiltonian $H=\dot{q}^A p_A - L$. Alternatively, one can start from the Hamiltonian formalism and look for functions $I=I(\tau,q,p)$ satisfying
\begin{equation} \label{Hamconslaw}
  \frac{dI}{d\tau} = \frac{\partial I}{\partial \tau} + \{I, H\} \approx 0,
\end{equation}
where with $\{\; ,\;\}$ we denote the usual Poisson brackets. The ``$\approx$'' symbol is used to take into account the possibility of the theory containing constraints; it denotes a weak equality in the theory of constrained systems \cite{Diracbook,Sund} and it offers a generalization of the usual definition of conserved charges that is applied in the Hamiltonian description of unconstrained systems \cite{Kuchar}. The weak equality means that the right hand side of \eqref{Hamconslaw} can either be directly zero or equal to a linear combination of constraints, which are first order quantities that have to vanish on mass shell.

An explicit symmetry breaking takes place, when the Lagrangian - and correspondingly the Hamiltonian - contain some parameter, say $\varepsilon$, whose value affects the symmetry group of the system. For example, if $\varepsilon=0$ we may have some group $A$ whose generators satisfy \eqref{symcrit}, while if $\varepsilon\neq 0$ the corresponding group changes to $B$ for which it usually holds that $B \subset A$. As it happens, both the symmetry criterion \eqref{symcrit} and the conservation law requirement \eqref{Hamconslaw}, define linear partial differential equations for the components of $X$ and the function $I$ respectively. If the Lagrangian and the Hamiltonian do not possess some pathology for the critical value $\varepsilon=0$, we can investigate the possibility that the number of the symmetries does not really change. In a sense, we may try to find solutions of either \eqref{symcrit} or \eqref{Hamconslaw} which continuously, through the parameter $\varepsilon$, connect cases $A$ and $B$.

In this work we present symmetry breaking examples where truly the number of symmetries does not change. The apparent symmetry breaking effect that we observe in this cases, is related to the fact that we consider partial solutions to \eqref{symcrit} or \eqref{Hamconslaw}. Usually, it is not realistic to obtain all possible vectors and functions satisfying the aforementioned equations, thus, we concentrate on a particular class of symmetries. However, what happens with the critical value of $\varepsilon$ is that some of the symmetries change the class to which they previously belonged, e.g. a point symmetry may become dynamical, or even nonlocal. This change is what we perceive as symmetry breaking when concentrating in the study of only one of these classes. We demonstrate this fact by examining closely two examples: firstly the motion of a free relativistic particle in a generic background metric and then the motion in a Bogoslovsky-Finsler space-time.

\section{The free relativistic particle}

For the motion of the free relativistic particle we use the Lagrangian in the einbein formalism
\begin{equation}\label{Lagfree}
   L = \frac{1}{2e} g_{\mu\nu} \dot{x}^\mu \dot{x}^\nu - e \frac{m^2}{2},
\end{equation}
where $e=e(\tau)$ is the einbein field. The sign conventions we use are compatible with space-times of Lorentzian signature $(-,+,+,+)$. The configuration space is spanned by the degrees of freedom $q^A =(e,x^\mu)$. Due to the Hessian matrix $\frac{\partial^2 L}{\partial \dot{q}^A \dot{q}^B}$ not being invertible, the Lagrangian $L$ is characterized as singular or constrained. Its corresponding Hamiltonian is obtained through the Dirac-Bergmann algorithm \cite{Dirac,AndBer} and it results in a linear combination of constraints
\begin{equation}\label{Hamree}
   H = \frac{e}{2} \mathcal{H} + u_e p_e,
\end{equation}
where $p_e \approx 0$ is the momentum for the einbein field $e$ and the primary constraint of the theory, while the
\begin{equation}\label{Hamcon}
  \mathcal{H} = p^\mu p_\mu + m^2 \approx 0
\end{equation}
is the secondary constraint (also called Hamiltonian or quadratic constraint) and $u_e$ is an arbitrary multiplier. The weak equality ``$\approx$'' in \eqref{Hamcon} basically means that the quantity is zero on mass shell, but its gradient in phase space is not. We refrain here from getting into details about constrained dynamics and refer to well known textbooks \cite{Diracbook,Sund} for further details.

Lagrangians of the form \eqref{Lagfree} and the corresponding Hamiltonian \eqref{Hamree} have a wide scope of application that goes outside the motion of a free relativistic particle. Cosmological minisuperspace Lagrangians can be mapped to the form of \eqref{Lagfree} by a reparametrization of the lapse function of the base manifold. In that case the $x^\mu$s are the configuration space variables related to the scale factors, the $g_{\mu\nu}$ is the minisuperspace metric and $e$ is proportional to the lapse function. This gives special importance to the study of the classical and quantum symmetries of such systems. For works on symmetries regarding cosmological Lagrangians see \cite{Cap,Andr,Dim3}. Studies for symmetries of geodesic motion in generic manifolds can be found in \cite{Andr2,Andr3}. There are also particular geometries which are of special physical interest, like plane gravitational waves, see \cite{Horv1,Horv2,Dim4}.

The parameter of this problem, which leads to an explicit symmetry breaking, is the mass $m$. It is well known that for null geodesics, when $m=0$, all conformal Killing vectors (CKVs), $\mathcal{L}_\xi g_{\mu\nu}=2\omega(x) g_{\mu\nu}$, are symmetries of the Lagrangian \eqref{Lagfree} and they generate linear in the momenta integrals of motion $I=\xi^\alpha p_\alpha$, since it can be easily seen that
\begin{equation}
  \frac{dI}{d\tau}=\{I,H\} = e \omega(x) p^\mu p_\mu = 0,
\end{equation}
with the last expression being zero due to $m$ vanishing in \eqref{Hamcon}, as it is expected for all null geodesics. However, in the massive case, linear in the momenta quantities are only conserved if $\xi$ is a pure Killing vector (KV) of the metric, i.e. $\mathcal{L}_\xi g_{\mu\nu}=0$. In that case we get directly $\{I,H\} =0$ without the use of the constraint \eqref{Hamcon}, since now $\omega(x)=0$. Thus, we see the constant $m$ being responsible for a symmetry breaking. The conformal Killing algebra of the $m=0$ case is broken to the subalgebra of the Killing vectors when $m\neq 0$. Let us investigate whether we can obtain a solution of \eqref{Hamconslaw} with a continuous transition from the one set of relative conserved quantities to the other through the parameter $m$. This is something which can truly be done, but in order to achieve it we must enhance the type of conserved quantities we are searching for.

To this extent we allow for the conserved quantity $I$ to have some explicit dependence on the parameter $\tau$ along the curve. Then, if $\xi$ denotes a conformal Killing vector of the space-time, $\mathcal{L}_\xi g_{\mu\nu}=2\omega(x) g_{\mu\nu}$, we can introduce the following quantity \cite{Dim1}:
\begin{equation}\label{nonloc}
  I(\tau,x,p) = \xi^\mu p_\mu + m^2 \int\!\! e(\tau) \omega\left(x(\tau)\right) d\tau,
\end{equation}
where the explicit dependence on $\tau$ is realized through the integral appearing in the right hand side of \eqref{nonloc}. The total derivative with respect to the parameter leads to
\begin{equation}
  \frac{dI}{d\tau} = \frac{\partial I}{\partial \tau} + \{\xi^\mu p_\mu,H\} = e \omega(x) \mathcal{H} \approx 0 .
\end{equation}
Thus, the $I$ of \eqref{nonloc} is conserved on the constrained surface and forms a nonlocal integral of the motion; nonlocal in the sense that it incorporates an integral of degrees of freedom of the problem.

Due to the conserved quantity \eqref{nonloc} being nonlocal, one would expect that some information about the trajectory needs to be known in order to calculate the explicit dependence on the variable $\tau$. This problem can be overcome at least for one such type of integral of motion due to the parametrization invariance corresponding to the freedom of choosing the parameter $\tau$. In the presence of a proper conformal Killing vector, the parameter along the curve can be taken so that the einbein field is equal to $e= \omega^{-1}$. Then, for the aforementioned vector, expression \eqref{nonloc} reduces to
 \begin{equation}\label{gfixedI}
  I = \xi^\mu p_\mu + m^2  \tau.
\end{equation}
This is exactly the type of conserved quantity that appears in the case of the affinely parametrized geodesics ($e=1$) when $\xi$ is a homothetic vector of the space-time metric ($\omega=1$).

Notice that \eqref{nonloc} re-establishes the continuity of the solution of \eqref{Hamconslaw} in what regards the value $m=0$. When the mass is zero we have $I=\xi^\mu p_\mu$ for all conformal Killing vectors, while for $m\neq 0$ we either have \eqref{nonloc} for proper conformal Killing vectors ($\omega\neq0$) or the purely linear expression $I=\xi^\mu p_\mu$ when ($\omega=0$), i.e. for Killing vectors. So, the symmetry breaking that we observe when $m\neq 0$ from CKVs to KVs happens only at the level of space-time vectors as generators of symmetries. The proper CKVs however, still contribute to symmetries, only that when $m\neq 0$, those do not correspond to pure space-time vectors but they additionally possess a nonlocal part and generate nonlocal integrals of motion. Thus, a new class of symmetries has taken the place of those which ``broke'' due to the mass being nonzero.

\section{Motion in a Bogoslovsky-Finsler space-time}

This case is motivated by a scenario of Lorentz violation due to some parameter $b$ in the Lagrangian and uses as the basic space-time that of a particular Finsler geometry. We remind that in a Finsler geometry the line element $ds^2= F(x,dx)^2$ is a homogeneous function of degree two in the differentials $dx^\mu$; the Riemannian sub-case being the one where it is purely quadratic. The Finsler metric is defined as
\begin{equation} \label{Finmet}
  G_{\mu\nu}(x,dx) = \frac{1}{2} \frac{\partial^2 F^2}{\partial(dx^\mu)\partial (dx^\nu)},
\end{equation}
where we notice that in the general case it also depends on the differentials $dx$. Particle dynamics in a Finsler geometrical context is often used to study dispersion relations emerging from Lorentz violation scenarios, see \cite{mech1,mech2,mech3,mech4}.

Let us consider the particular geometry given by the Bogoslovsky-Finsler line element \cite{Bogo1,Bogo2}
\begin{equation} \label{Bogolineel}
  ds^2 = \eta_{\mu\nu}dx^\mu dx^\nu \left[\frac{\left(\ell_\mu dx^\mu\right)^{2}}{-\eta_{\mu\nu}dx^\mu dx^\nu}\right]^b ,
\end{equation}
with $\eta_{\mu\nu}$ being the flat space-time metric and $\ell$ a covariantly constant, null, future directed Killing vector and $0\leq b<1$. If $b\neq 0$ the geometry is Finslerian, while when  $b$ is zero, the typical pseudo-Riemannian line element of Special Relativity is obtained. Experimental constraints assign a very small value to the parameter $b<10^{-26}$ \cite{GiGoPo}. More in this modification of Special Relativity can be found in \cite{Bogobook}; for a study on the integrability of the geodesics when the metric is that of a pp-wave see \cite{ppBogo}.

In a similar manner to \eqref{Lagfree}, the motion of a free particle in such a space-time is given by
\begin{equation} \label{genLagF}
  L = \frac{1}{2 e} G_{\mu\nu}(x,\dot{x}) \dot{x}^\mu \dot{x}^\nu - e \frac{m^2}{2}.
\end{equation}
It is more convenient to work in light cone coordinates, $x^\mu=(v,u,\x,\y)$, we thus consider
\begin{equation}
  \eta_{\mu\nu} dx^\mu dx^\nu = 2 du dv + \delta_{ij} dx^i dx^j,
\end{equation}
while for the vector $\ell$ we write $\ell =\ell^\mu \partial_\mu= \partial_{v}$. As a result, the Lagrangian function finally reads
\begin{equation} \label{LagF}
  L = -\frac{1}{2 e} \dot{u}^{2b} \left(-\dot{x}^\mu \dot{x}_\mu \right)^{1-b} -e \frac{m^2}{2} .
\end{equation}

If we start from the infinitesimal criterion of invariance \eqref{symcrit} and consider pure point symmetries, which means that for the components of the generator \eqref{symvecgen} we have $\chi=\chi(\tau,q)$ and $\xi^A=\xi^A(\tau,q)$, where $q^A=(x^\mu,e)=(v,u,\x,\y,e)$, then, upon solving the resulting set of equations, we arrive at a branching point where we need to decide whether the parameter $b$ is zero or not. Depending on this choice the solution space changes. What is common in both cases is the resulting parametrization invariance generator, which we leave outside our considerations, and the fact that the vector $\xi =\xi^\mu(x) \frac{\partial}{\partial x^\mu}$ is a pure space-time vector.

The value $b=0$ yields the $\xi$ as an element of the Poincar\'e, $\mathfrak{iso}(3,1)$, algebra. That is, $\xi$ is any linear combination of the following vectors involving translations, rotations and boosts
\begin{subequations} \label{Poiten}
  \begin{align} \label{Poiseven}
    T_\mu = \partial_{\mu},& \quad B_{ij} = x_i \partial_{j} - x_j  \partial_{i} , \quad B_{v i} = u \partial_{i} - x_i \partial_{v} \\
    & B_{u v} = v \partial_{v} - u \partial_u , \quad B_{ui} = v \partial_i -x_i \partial_u .   \label{Poithree}
  \end{align}
\end{subequations}
In the above expressions the indexes $u$, $v$ are used to denote the relative directions, while the $i$, $j$ take the values $1,2$ and denote the $\x$, $\y$ directions respectively.

The result \eqref{Poiten} is in accordance to what we expect from the theory since $b=0$ reduces \eqref{LagF} to the free relativistic particle Lagrangian in Minkowski space. However, the $b\neq 0 $ case opens a different branch and the resulting solution for $\xi$ consists just of the seven vectors of \eqref{Poiseven} plus a new vector
\begin{equation} \label{Nbsym}
  N_b = (1+b) v \partial_{v} + (b-1) u \partial_u +  b x^i \partial_i .
\end{equation}
These eight vectors form the $\mathfrak{disim}_b(2)$ algebra which is a deformation \cite{GiGoPo} of the $\mathfrak{isim}(2)$ algebra of Very Special Relativity (VSR) \cite{Cohen}. The main idea behind VSR is to maintain just a subgroup of the Lorentz group as the basic symmetry of nature. This has motivated several authors to seek its implications \cite{Alfaro,Lee,Bufalo}. Here, we are interested in the deformed $\mathfrak{disim}_b(2)$ algebra whose nontrivial commutators are
\begin{equation}
  \begin{split}
   & [N_b, T_v]= -(1+b)T_v, \quad [N_b, T_u] = (1-b) T_u, \\
   & [N_b,T_i]=-b T_i, \quad [N_b, B_{v i}] = -B_{v i} \\
   & [T_u,B_{vi}]= T_i, \quad [T_i,B_{12}] = \epsilon_{ij}T_j, \\
   & [T_i,B_{vj}]= - \delta_{ij} p_v, \quad [B_{vi},B_{12}]=\epsilon_{ij}B_{vj} ,
  \end{split}
\end{equation}
where $\epsilon_{ij}$ is the antisymmetric tensor with $\epsilon_{12}=1$. We thus see that for the value $b\neq0$ we have a Lorentz violation due to the ``disappearance'' of the three symmetries \eqref{Poithree} that emerge in the $b=0$ case. Note that at the limit $b\rightarrow 0$, the vector $N_b$ of \eqref{Nbsym} becomes the $B_{u v}$ of \eqref{Poithree}.

We try to find a more general solution to \eqref{symcrit} incorporating a smooth transition through the critical value $b=0$. Once more, the three symmetries \eqref{Poithree} do not simply disappear. What happens is that they cease to be point symmetries corresponding to space-time vectors. Truly, if we search for a more generalized type of symmetries, i.e. generators of contact transformations, $\xi^\mu = \xi^\mu(x,\dot{x})$, then we can see that in place of \eqref{Poithree}, when $b\neq 0$, the following symmetries of the Lagrangian \eqref{LagF} emerge \cite{Dim2}:
\begin{subequations} \label{symvecs}
\begin{align} \label{symvecs1}
  \xi_{uv} & = - u \partial_u + \left(v+ \frac{b}{1-b} \frac{\dot{x}^\mu \dot{x}_\mu}{\dot{u}^2} u  \right) \partial_v \\
  \xi_{ui} & = - x_i \partial_u + \frac{b}{1-b}\frac{\dot{x}^\mu \dot{x}_\mu}{\dot{u}^2}  x_i  \partial_v + v \partial_i \; .
\end{align}
\end{subequations}
Notice how once more the continuity is restored since, when $b=0$, the $\xi_{uv}$, $\xi_{ui}$ become the $B_{uv}$, $B_{ui}$ respectively. Thus, for a generic nonzero $b$ we have the higher order symmetries \eqref{symvecs}, which reduce to space-time vectors when $b=0$. The variational symmetries \eqref{symvecs} give rise to the Noether charges
\begin{subequations}\label{extraI}
\begin{align}
  I_{uv} & =  v p_v - u p_u + \frac{b}{1+b} u \frac{p_\mu p^\mu}{p_v}   \\
  I_{ui} & = v p_i - x_i p_u  + \frac{b}{1+b} x_i \frac{p_\mu p^\mu}{p_v} .
\end{align}
\end{subequations}
which we have expressed in phase space coordinates. It is easy to verify that their Poisson brackets with the corresponding Hamiltonian constraint, given by
\begin{equation}\label{HamBg}
  \mathcal{H} = -\frac{(1-b)^{b-1}}{(1+b)^{1+b}} p_v^{-2 b} \left(-p_\mu p^\mu \right)^{1+b} + m^2 \approx 0,
\end{equation}
are zero,  $\{I_{uv},H\}=0=\{I_{ui},H\}$, where the total Hamiltonian $H$ is again given by \eqref{Hamree}. The Lorentz violation of the $b\neq 0$ case, gives rise to the appearance of a different type of symmetries \eqref{symvecs} and conserved charges \eqref{extraI}, in which we see the modification that the nonzero parameter $b$ induces to the corresponding boosts of the Minkowski case.

However, apart from $b$ we also have the parameter $m$ in the Lagrangian, which according to the previous section is also responsible for breaking symmetries related to conformal Killing vectors. It is easily verified that the following five vectors
\begin{subequations} \label{confb}
\begin{align}
   \xi_D =& \left(M_b(\dot{x})^2 u + 2 v\right) \partial_v + x^i \partial_i \\
   \xi_K =& u^2 \partial_u + \frac{1}{2} \left(\frac{1+b}{1-b} M_b(\dot{x})^2 u^2 - \mathbf{x}^2 \right) \partial_v + u x^i \partial_i \\
   \xi_{C_1} =& \frac{\mathbf{x}^2}{2}  \partial_u - \frac{1}{4} \left[ M_b(\dot{x})^4 u^2 + M_b(\dot{x})^2 \left( 4 u v - \frac{1-b}{1+b}\mathbf{x}^2 \right)+ 4 v^2 \right] \partial_v    - \left(\frac{M_b(\dot{x})^2}{2} u + v\right)x^i \partial_i \\
   \xi_{C_2}^j =& u x^j \partial_u + \left(\frac{1}{1-b}M_b(\dot{x})^2 u +v \right)x^j\partial_v - \frac{1}{2} \left(M_b(\dot{x})^2 u^2 +2 u v +\mathbf{x}^2 \right) \partial_j + x^j x^i \partial_i,
\end{align}
\end{subequations}
where
\begin{equation}
  M_b(\dot{x})^2 = -\frac{\dot{x}^\mu\dot{x}_\mu}{\dot{u}^2}
\end{equation}
and $\mathbf{x}^2= \x^2+\y^2$ are higher order symmetries of \eqref{LagF}. For more on higher order symmetries in dynamical systems see a recent study in  \cite{Mits}. Note that the function $M_b(\dot{x})$ is constant on mass shell
\begin{equation}
  M_b^2 \overset{!}{=} \left[\frac{(1-b)^2 m^2}{p_v^2}\right]^{\frac{1}{1+b}} ,
\end{equation}
with $p_v$ also a constant since all translations are symmetries of the problem, see \eqref{Poiseven} for the vectors that are symmetries still when $b\neq 0$. Hence, either set of vectors, \eqref{symvecs} or \eqref{confb} can be seen on mass shell as $m$ and $b$ distorted space-time vectors generating disformal transformations \cite{Dim2}. For details on disformal transformations and applications see \cite{Bekenstein,Lobo1,Lobo2}.

When the mass $m$ is zero, there is no contribution from $M_b$ and the vectors \eqref{confb} reduce to the usual expressions for the five proper conformal Killing vectors of Minkowski space as given in light cone coordinates. It is no difficult task to write once more the integrals of motion of the form $I=\xi^\mu(x,p) p_\mu$ that contain rational functions in the momenta and which of course commute with the Hamiltonian of the system.

Interestingly enough the additional space-time vector $N_b$ of \eqref{Nbsym}, that appears as an extra symmetry in the $b\neq 0$ case, is given by the linear combination
\begin{equation}
  N_b = b \xi_D + \left(1-b\right) \xi_{uv},
\end{equation}
i.e. the use of the distorted homothecy plus the distorted boost from \eqref{symvecs1}. We observe that, the circle is once more completed and the connection to the results of the $m=0=b$ is restored in a continuous manner.

\section{Conclusions}

We examined two cases where the explicit symmetry breaking leads to the generation of different types of symmetries out of distorting appropriately those which where broken. The first example had to do with the motion of a relativistic particle in a generally curved space. The existence of the mass breaks the symmetry that the null geodesics have under the conformal algebra of the metric. However, we saw how the proper conformal Killing vectors contribute in nonlocal integrals of motion exactly because of the nonzero mass. The second example regards Lorentz violation through the motion in the Bogoslovsky-Finsler space-time. The violation parameter breaks the symmetry of three of the Killing vectors of Minkowski space. Again, with appropriate distortions these vectors can lead to new symmetries and their corresponding conserved charges; only this time, these are higher order symmetries and the integrals of motion are rational functions in the momenta. The mass of the particle has a similar effect, and - as in the pseudo-Riemannian case - it also affects the proper conformal Killing vectors.

We argue that this is not an isolated effect regarding these particular examples. It is something which may be generalized in several other cases as well. The  linearity of the symmetry criterion \eqref{symcrit} and of the corresponding condition that conserved charges need to satisfy (see eq. \eqref{Hamconslaw}) contributes to this hypothesis. Together with the fact that, in physical theories, the involved functions usually satisfy certain smoothness conditions. Nevertheless, further examples have to be studied in this regard which may lead to interesting new results regarding additional symmetries and conservation laws, especially in field theories.

\section*{Acknowledgements}
N. D. acknowledges the support of the Fundamental Research Funds for the Central Universities, Sichuan University Full-time Postdoctoral Research and Development Fund No. 2021SCU12117.

\end{document}